\documentclass{notemat}
\usepackage{amsmath}
\usepackage{amssymb}
\usepackage{nmmacro}
\usepackage[fleqn, intlimits]{empheq}
\usepackage{cancel}
\chardef\bslchar=`\\ 
\makeatletter
\newcommand{\addbslash}{\expandafter\@addbslash\string}
\def\@addbslash#1{\bslchar\@nobslash#1}
\newcommand{\nobslash}{\expandafter\@nobslash\string}
\def\@nobslash#1{\ifnum`#1=\bslchar\else#1\fi}
\newcommand{\ntt}{\normalfont\ttfamily}
\def\@boxorbreak{\leavevmode
  \ifmmode\hbox\else\ifdim\lastskip=\z@\penalty9999 \fi\fi}
\DeclareRobustCommand{\cs}[1]{\@boxorbreak{\ntt\addbslash#1\@empty}}
\makeatother
\def\dlap{\tilde{\nabla}^2}

\begin{document}
\begin{article}
\begin{opening}
  \title{On a class of rational matrices and interpolating polynomials related to the discrete Laplace operator}
\author{Pierpaolo Vivo\thanks{This work is supported by a Marie Curie Host Fellowship for Early Stage Researchers Training (NET-ACE project).}}
  \institute{School of Information Systems, Computing \& Mathematics\\ Brunel
  University\\
   Uxbridge, Middlesex (United Kingdom)\\
    \email{pierpaolo.vivo@brunel.ac.uk}}
\author{Mario Casartelli}
 \institute{Dipartimento di Fisica, Universit\'a degli Studi di
 Parma\\
 Parco Area delle Scienze 7a, 43100 Parma (Italy)\\INFN, Gruppo
 collegato di Parma\\CNR-INFM, Parma\\
    \email{casartelli@fis.unipr.it}}
\author{Luca Dall'Asta}
 \institute{Abdus Salam International Center for Theoretical Physics,\\ Strada
  Costiera 11, 34014, Trieste (Italy)\\
    \email{dallasta@ictp.it}}
\author{Alessandro Vezzani}
 \institute{Dipartimento di Fisica, Universit\'a degli Studi di
 Parma\\
 Parco Area delle Scienze 7a, 43100 Parma (Italy)\\CNR-INFM, Parma\\
    \email{Alessandro.Vezzani@fis.unipr.it}}
  \runningauthor{P. Vivo, M. Casartelli, L. Dall'Asta and A. Vezzani}
  \runningtitle{A class of discrete harmonic rational matrices}
\begin{abstract}
Let $\dlap$ be the discrete Laplace operator acting on functions
(or rational matrices) $f:\mathbf{Q}_L\rightarrow\mathbb{Q}$,
where $\mathbf{Q}_L$ is the two dimensional lattice of size $L$
embedded in $\mathbb{Z}_2$. Consider a rational $L\times L$ matrix
$\mathcal{H}$, whose inner entries $\mathcal{H}_{ij}$ satisfy
$\dlap\mathcal{H}_{ij}=0$. The matrix $\mathcal{H}$ is thus the
classical finite difference five-points approximation of the
Laplace operator in two variables. We give a constructive proof
that $\mathcal{H}$ is the restriction to $\mathbf{Q}_L$ of a
discrete harmonic polynomial in two variables for any $L>2$. This
result proves a conjecture formulated in the context of
deterministic fixed-energy sandpile models in statistical
mechanics.
\end{abstract}


\end{opening}

\section*{Introduction and Motivation}
An interesting class $\mathcal{M}_L$ of $L\times L$ matrices
$\mathcal{H}$ with rational entries and a related vector space of
polynomials in two variables arise in some theoretical physics
models, the so-called \emph{deterministic fixed-energy
sandpiles} (\textsf{DFES}) with Bak-Tang-Wiesenfeld (\textsf{BTW})
toppling rule \cite{Sandpile}.

Introduced for the first time in \cite{Dickman} by imposing a
global energy conservation constraint on its dissipative
counterpart \cite{Bak}, \textsf{DFES} is a deterministic cellular
automaton, in which two-dimensional configurations (represented by
square matrices with integer elements $z_{ij}(t)$) evolve in discrete time steps $t$
according to a precise parallel updating rule.

The main feature of \textsf{DFES} is that, in contrast with the dissipative model,
only a small part of an a priori huge configuration space is dynamically explored,
and the system enters a periodic orbit after a surprisingly short transient. This is a
clear indication of the existence of many \emph{hidden
conservation laws} (\textsf{HCL}) which split the wide
configuration space into dynamically intransitive, and thus much
smaller subspaces \cite{Sandpile}.

Few of those \textsf{HCL} were identified in a non-systematic way
in \cite{Bagnoli} and can be represented in the form:
\begin{equation}\label{HCL}
  \Phi_L[f](t)=\left[\sum_{i,j}f(i,j)z_{ij}(t)\right]\textrm{ mod
  }L
\end{equation}
where the sum runs over the integer coordinates of the
two-dimensional $L\times L$ lattice sites, $z_{ij}(t)$ is the
integer value taken by the entry $(i,j)$ at time $t$ and $f(i,j)$
is a $L\times L$ matrix with rational entries. The interest is
then in characterizing the \emph{generating functions}
(\textsf{GF}) of \textsf{HCL}, i.e. the class of inequivalent
matrices $f$ such that $\Phi_L[f](t)$ is a \textsf{HCL}
($\Phi_L[f](t+1)=\Phi_L[f](t)$ for all $t$).

Bagnoli et al. \cite{Bagnoli} gave the following three
\textsf{GF}: $f_1=i$, $f_2=j$ and $f_3=i^2-j^2$. An intriguing
observation is that, when thought as functions on the whole
$\mathbb{R}^2$ ($f(x,y):\mathbb{R}^2\mapsto\mathbb{R}$), those
three \textsf{GF} belong to a special vector space of polynomials
in two variables, which we call \emph{discrete harmonic
polynomials} (see Def. \ref{Dhpoly}). It is then appealing to
conjecture that this should be a general feature of any
\textsf{GF} of a \textsf{HCL}.

In fact, an exhaustive characterization of \textsf{GF} has been
given in \cite{Sandpile} from a completely
different perspective, i.e. \emph{without any reference to
polynomials}, but working simply on the matrix
representation of those \textsf{GF}.

It was proven in \cite{Sandpile} that a functional of the form
\eqref{HCL} is a \textsf{HCL} if and only if its \textsf{GF} is a
\emph{inner-harmonic matrix of size $L$} (see Def.
\ref{InnHarmMat})\footnote{In appendix B of \cite{Sandpile} the
necessary and sufficient condition is expressed in terms of
K-harmonicity, and strictly speaking this is not equivalent to
inner-harmonicity. However, it can be proved that for every
K-harmonic function there exists an inner-harmonic function which
belongs to the same equivalence class, i.e. generates an
equivalent \textsf{HCL}. Thus, it is not restrictive to work with
inner-harmonic matrices, as we will do from now on.}.

The purpose of this paper is to provide a rigorous link between
the exact characterization of \textsf{HCL} in terms of matrices
\cite{Sandpile} and the conjectured polynomial form for any
\textsf{GF}. More precisely, we will prove that every
inner-harmonic matrix of size $L$ (i.e. any \textsf{GF} of a
\textsf{HCL} in the sandpile context) can be represented (non
uniquely) as the restriction to the two-dimensional discrete
lattice of a discrete harmonic polynomial in two variables .

The paper is organized as follows. In Section \ref{Definitions} we
set up notations and basic definitions, providing in particular
the notions of i) \emph{inner-harmonic matrix of size $L$} (Def.
\ref{InnHarmMat}), in terms of the well-known five-points formula
for the discretization of the Laplace operator on a 2d lattice,
and ii) \emph{discrete harmonic polynomial} (Def. \ref{Dhpoly}).
In Section \ref{Main Results}, we enunciate the main theorem and
provide the algorithmic procedure for finding the discrete
harmonic polynomial which interpolates any given inner-harmonic
matrix of size $L\geq 3$. In the same section, we provide a
stepwise example of application, together with pointers to
subsequent lemmas needed for the proof. Section \ref{FinalRemarks}
is devoted to conclusive remarks and hints for future works, while
a basis of discrete harmonic polynomials up to degree $9$ is given
in the Appendix.

\section{Definitions}\label{Definitions}
We define $\mathbf{Q}_L$ as the two dimensional lattice embedded
in $\mathbb{Z}_2$, i.e.:
\begin{equation}
\mathbf{Q}_L=\{(i,j)\in\mathbb{Z}_2|0\leq i,j\leq L-1\}
\end{equation}

\begin{Definition}
The \emph{inner sublattice} $\mathbf{Q}^\dag_L$ of $\mathbf{Q}_L$ is
the set:
\begin{equation}
\mathbf{Q}^\dag_L=\{(i,j)\in\mathbf{Q}_L|1\leq i,j\leq L-2\}
\end{equation}
\end{Definition}

The discrete Laplace operator is defined as the classical finite
difference five-points second order formula for the approximation
of the Laplace operator:
\begin{Definition}
Let $f:\mathbf{Q}_L\rightarrow \mathbb{Q}$. The discrete laplace
operator $\dlap$ acts on $f$ as:
\begin{equation}\label{dlap}
(\dlap f^\dag)(i,j)=4f(i,j)-f(i-1,j)-f(i+1,j)-f(i,j-1)-f(i,j+1)
\end{equation}
where $f^\dag:=f|_{\mathbf{Q}^\dag_L}$.

The generalization to functions
$f:\mathbb{R}^2\rightarrow\mathbb{R}$ is straightforward (consider
$f\equiv f^\dag$ in this case).
\end{Definition}

\begin{Definition}
Let $\mathcal{M}_L$ be the set of rational $L\times L$ matrices and
$\mathfrak{F}=\{f|f:\mathbf{Q}_L\rightarrow \mathbb{Q}\}$. We define
the invertible map $\Psi:\mathfrak{F}\rightarrow \mathcal{M}_L$
(\emph{$L$-correspondence}) through the following:
\begin{equation}\label{correspondence}
 \Psi(h):=\mathcal{H}
 \end{equation}
 where $h(-1+j,-i+L):=\mathcal{H}_{i,j}$.

 Through $\Psi$, the lower left corner of $\mathcal{H}$ is mapped
 to the point $(0,0)$.
 \end{Definition}
\begin{Definition}\label{InnHarmMat}
A $L\times L$ rational matrix $\mathcal{H}_1$ is called
\emph{inner-harmonic matrix of size $L$} ($L>2$) if the following
property holds ($h_1=\Psi^{-1}(\mathcal{H}_1)$):
\end{Definition}
\begin{equation}\label{InnHarmMatCondition}
(\dlap h_1^\dag)(i,j)=0
\end{equation}
as in the following example, where we restrict for simplicity to
integer entries:
\begin{equation}
\tilde{\mathcal{H}}=\left(
\begin{array}{ccccccc}
  2  &  0       &  0        &  1        &  0        &  1        &  2 \\
  0  &  2       &  1        &  2        &  0        &  2        &  1 \\
  1  &  7       &  0        &  6        &  -4       &  6        &  2 \\
  1  &  25      &  -14      &  26       &  -28      &  24       &  1 \\
  2  &  106     &  -107     &  140      &  -158     &  117      &  2 \\
  2  &  504     &  -660     &  799      &  -861     &  600      &  1 \\
  1  &  2568    &  -3836    &  4577     &  -4685    &  3143     &  0 \\
\end{array}
 \right)
\end{equation}
\begin{Definition}\label{Dhpoly}
A polynomial $P(x,y)$ is called \emph{discrete harmonic polynomial}
if $(\dlap P)(x,y)=0\quad\forall (x,y)\in\mathbb{R}^2$.
\end{Definition}
Examples of discrete harmonic polynomials are
$P_1(x,y)=x^2-y^2,P_2(x,y)=x^3-3xy^2,P_3(x,y)=xy$.

The set of discrete harmonic polynomials of degree $g$ will be
denoted as $\mathbb{D}^\star_g$.

\begin{Definition}
We say that a polynomial $P(x,y)$ \emph{interpolates} a $L\times
L$ matrix $\mathcal{H}$ if $P(i,j)=h(i,j)$, where
$\Psi(h)=\mathcal{H}$. In this case, we write $P\doteq
\mathcal{H}$.
\end{Definition}
Note that:
\begin{Remark}
  Discrete harmonic polynomials are generally \emph{not} harmonic in
  $\mathbb{R}^2$, i.e. solutions of the continuum Laplace equation
  $\nabla^2 P=0$. Generally speaking, every polynomial $\mathcal{P}(x,y)$ in two variables belongs to
  one of the following classes:
\begin{itemize}\label{ClassesPolynomial}
  \item $\mathcal{P}(x,y)$ is neither harmonic nor discrete
  harmonic. Example: $\mathcal{P}(x,y)=x^3+y^3$
  \item $\mathcal{P}(x,y)$ is harmonic but not discrete
  harmonic. Example: $\mathcal{P}(x,y)=x^4-6 x^2 y^2+y^4$
  \item $\mathcal{P}(x,y)$ is discrete harmonic but not
  harmonic. Example: $\mathcal{P}(x,y)=x^4-2 x^2-6 x^2 y^2+y^4$
  \item $\mathcal{P}(x,y)$ is both harmonic and discrete harmonic.
  Example: $\mathcal{P}(x,y)=xy$
\end{itemize}

\end{Remark}
\begin{Remark}
  Given a discrete harmonic polynomial $P(x,y)$, it obviously
  interpolates an
  inner-harmonic matrix $\mathcal{H}_L$ on $\mathbf{Q}_L$.
  For example, the polynomial $P(x,y) =x^3-3xy^2$
  interpolates the following matrix on $\mathbf{Q}_7$:
  \begin{equation}\label{Restriction}
    \mathcal{H}_7 =\left(
\begin{array}{ccccccc}
  216 &  198  &  144  &  54   &  -72  &  -234  &  -432 \\
  125 &  110  &  65   &  -10  &  -115 &  -250  &  -415 \\
  64  &  52   &  16   &  -44  &  -128 &  -236  &  -368 \\
  27  &  18   &  -9   &  -54  &  -117 &  -198  &  -297 \\
  8   &  2    &  -16  &  -46  &  -88  &  -142  &  -208 \\
  1   &  -2   &  -11  &  -26  &  -47  &  -74   &  -107 \\
  0   &  0    &  0    &  0    &  0    &  0     &  0\\
\end{array}
 \right)
  \end{equation}
  \end{Remark}

The converse is not trivial for any $L>2$: while it is
straightforward to find an interpolating polynomial $\Phi(x,y)$
for any given inner-harmonic matrix through any of the known
Polynomial Interpolation formulas in two variables
\cite{interpol}, the resulting $\Phi$ is generally \emph{not}
discrete harmonic in $\mathbb{R}^2$ (and incidentally neither
harmonic). This can be seen easily by referring to the widely used
Bilinear Interpolation formula (see e.g. \cite{Wiki}), the
extension to the two-dimensional lattice of the well-known
Lagrange interpolation formula in 1d:
\begin{equation}\label{Lagrange}
   \Phi(x,y)=\sum_{h,k}z_{hk}\prod_{\stackrel{j=0}{j\neq h}}^{L-1}
   \frac{x-j}{h-j}
   \prod_{\stackrel{r=0}{r\neq k}}^{L-1}
   \frac{y-r}{k-r}
\end{equation}
where $z_{hk}=\mathcal{H}_{h,k}$, the sum runs over the sites of
the matrix and the products over rows and columns respectively.
Note that $\mathrm{deg}(\Phi)=2(L-1)$.

It is then possible to interpolate the following simple
inner-harmonic matrix of size $L=4$:
\begin{equation}\label{Interpol}
    \mathcal{H}_4 =\left(
\begin{array}{ccccccc}
  27  &  18   &  -9   &  -54 \\
  8   &  2    &  -16  &  -46 \\
  1   &  -2   &  -11  &  -26  \\
  -3   &  0    &  0    &  0  \\
\end{array}
 \right)
  \end{equation}
The bilinear interpolating polynomial is the following:
\begin{align}\label{phi4interpol}
\nonumber\Phi_{\mathcal{H}_4}(x,y) &= -3+\frac{11}{2}x-3 x^2
+\frac{1}{2}x^3+\frac{11}{2}y-\frac{121}{12}xy+\frac{5}{2}x^2 y
-\frac{11}{12}x^3 y+\\
&-3 y^2 +\frac{11}{2}x y^2 -3 x^2 y^2+\frac{1}{2}x^3 y^2
+\frac{3}{2}y^3 -\frac{11}{12}x y^3+\frac{1}{2}x^2
y^3-\frac{1}{12}x^3 y^3
\end{align}
and a straightforward calculation yields
$\dlap(\Phi_{\mathcal{H}_4})(x,y)\neq 0$ in $\mathbb{R}^2$.

In the following section, we shall provide the enunciation of
the main result, a stepwise example of application of the
algorithm, and a constructive proof of the main theorem.

\section{Interpolation by discrete harmonic polynomials: main result and algorithm}\label{Main Results}
We enunciate our main result:
\begin{Theorem}\label{main}
Let $\mathcal{H}$ be an inner-harmonic matrix of size $L>2$. There
exists a discrete harmonic polynomial $P(x,y)$ of degree less or
equal to $2(L-1)$ such that $P$ interpolates $\mathcal{H}$ on
$\mathbf{Q}_L$.
\end{Theorem}
Before getting to the technical points, it is informative to
provide an example of how our algorithmic procedure roughly works.

Let us consider the inner-harmonic matrix
$\mathcal{H}:=\mathcal{H}_4$ in \eqref{Interpol}.\\
\\
\verb"First step:"\\
Isolate the lower left $(3\times 3)$ minor
$\mathcal{H}^{(1)}\subset
 \mathcal{H}$:
\begin{equation}\label{Minor}
    \mathcal{H}^{(1)} =\left(
\begin{array}{ccccccc}
  8   &  2    &  -16  \\
  1   &  -2   &  -11  \\
  -3   &  0    &  0    \\
\end{array}
 \right)
\end{equation}\\
\\
\verb"Second step:"\\
Apply Lemma \ref{3x3} and find a discrete harmonic
polynomial\footnote{Note that this polynomial does
NOT coincide with the bilinear interpolating polynomial we would
get for the same matrix.} $P^{(1)}(x,y)\doteq\mathcal{H}^{(1)}$:
\begin{align}
\nonumber P^{(1)}(x,y) &=-3+\frac{15}{4}x-\frac{1}{8}x^2-
\frac{3}{4}x^3+\frac{1}{8}x^4+\frac{15}{4}y+\\
&-\frac{27}{4}xy-\frac{3}{4}x^2 y-\frac{1}{8}y^2+\frac{9}{4} x y^2
-\frac{3}{4}x^2 y^2
+\frac{1}{4}y^3+\frac{1}{8}y^4
\end{align}\\
\\
\verb"Third step:"\\
Evaluate $P^{(1)}(x,y)$ on the lattice $\mathbf{Q}_L\equiv
\mathbf{Q}_4$, obtaining the matrix $\hat{\mathcal{H}}_4$:
\begin{equation}\label{Hhat4}
    \hat{\mathcal{H}}_4 =\left(
\begin{array}{ccccccc}
24 & \boxed{18} & -9 & -57\\
  8   &  2    &  -16 & -46 \\
  1   &  -2   &  -11  & \boxed{-26}\\
  -3   &  0    &  0   & -3 \\
\end{array}
 \right)
 \end{equation}
Note that i) $\hat{\mathcal{H}}_4\neq \mathcal{H}$ ii) the sites
$(1,3)=18$ and $(3,1)=-26$ are uniquely determined by the discrete
harmonicity requirement and thus coincide in the two matrices.\\
\\
\verb"Fourth step:"\\
In order to amend the other mismatching entries along the border,
compute the four \textsf{(L)-Polynomials} (Lemma
\ref{ZLpolynomials}):
\begin{align}
\nonumber \xi_1(x,y) &=384 x -656 x^2+375 x^3-65 x^4-3 x^5+x^6-516 y+332 xy+\\
\nonumber &+465 x^2 y-440 x^3 y+105 x^4 y-6 x^5 y + 776 y^2-1095 x y^2+\\
\nonumber &+360 x^2 y^2 + 30 x^3 y^2 -15 x^4 y^2 -225 y^3+460 x
y^3-210 x^2y^3+\\
&+20 x^3 y^3-55 y^4-15 x y^4+15 x^2 y^4 + 21 y^5-6 x y^5-y^6\\
\nonumber \xi_2(x,y) &=240 x-386 x^2+135 x^3+25 x^4-15 x^5+x^6-168 y-152 x y+\\
\nonumber &+555 x^2 y-280 x^3 y+15 x^4 y+6 x^5 y+326 y^2-255 x
y^2+\\
\nonumber &-180 x^2 y^2+150 x^3 y^2-15 x^4 y^2-195 y^3+260 x
y^3-30 x^2 y^3+\\
&-20 x^3 y^3 + 35 y^4-75 x y^4 +15 x^2 y^4 +3 y^5+6 x
y^5-y^6\\
\nonumber \xi_3(x,y) &=516 x -776 x^2+225 x^3+55 x^4-21 x^5+ x^6-348 y -332 x y+\\
\nonumber &+1095 x^2 y-460 x^3 y+15 x^4 y+6 x^5 y+656 y^2-465 x y^2+\\
\nonumber &-360 x^2 y^2+210 x^3 y^2-15 x^4 y^2-375 y^3+440 x
y^3-30 x^2
y^3+\\
&-20 x^3 y^3+65 y^4-105 x y^4+15 x^2 y^4+3 y^5+6 x y^5-y^6\\
 \nonumber\xi_4(x,y) &=1644 x -2852 x^2+1305 x^3-35 x^4-69 x^5+7
 x^6+\\
 \nonumber &-1644 y+3225 x^2 y-2130 x^3 y+345 x^4 y+2852 y^2+\\
 \nonumber &-3225 x y^2+690 x^3 y^2-105 x^4 y^2-1305 y^3+2130 x y^3+\\
 &-690 x^2 y^3+35 y^4-345 x y^4+105 x^2 y^4 + 69 y^5 -7
 y^6
\end{align}
Those $\xi_k$ have the remarkable properties to be i) discrete
harmonic in $\mathbb{R}^2$ ii) almost everywhere $0$ on
$\mathbf{Q}_L$, except one single entry (two for $\xi_4$). In
particular, $\xi_k\doteq  \hat{\mathcal{\xi}}_k$, where:
\begin{equation}\label{xi1mat}
    \hat{\mathcal{\xi}}_1 =\left(
\begin{array}{ccccccc}
\gamma_1 & 0 & 0 & 0\\
  0  &  0    &  0 & 0 \\
  0   &  0   &  0  & 0\\
  0   &  0    &  0   & 0 \\
\end{array}
 \right)
 \end{equation}
\begin{equation}\label{xi2mat}
    \hat{\mathcal{\xi}}_2 =\left(
\begin{array}{ccccccc}
 0 & 0 & 0 & \gamma_2\\
  0  &  0    &  0 & 0 \\
  0   &  0   &  0  & 0\\
  0   &  0    &  0   & 0 \\
\end{array}
 \right)
 \end{equation}
 \begin{equation}\label{xi3mat}
    \hat{\mathcal{\xi}}_3 =\left(
\begin{array}{ccccccc}
0 & 0 & 0 & 0\\
  0  &  0    &  0 & 0 \\
  0   &  0   &  0  & 0\\
  0   &  0    &  0   & \gamma_3 \\
\end{array}
 \right)
 \end{equation}
 \begin{equation}\label{xi4mat}
    \hat{\mathcal{\xi}}_4 =\left(
\begin{array}{ccccccc}
0 & 0 & \gamma_4 & 0\\
  0  &  0    &  0 & -\gamma_4 \\
  0   &  0   &  0  & 0\\
  0   &  0    &  0   & 0 \\
\end{array}
 \right)
 \end{equation}
where, for the particular choice of the basis polynomials used to
build up the $\xi(x,y)$ (see Lemma \ref{ZLpolynomials} for
details), we have
$(\gamma_1,\gamma_2,\gamma_3,\gamma_4)=(-720,-720,720,-720)$.\\
\\
\verb"Fifth step:"\\
Define the sought interpolating polynomial $P(x,y)$ for
$\mathcal{H}$ as:
\begin{equation}\label{WantedInterpolating}
P(x,y)=P^{(1)}(x,y)+\sum_{k=1}^4 z_k \xi_k(x,y)
\end{equation}
where $z_k$ are parameters to be determined, and compute $P(x,y)$
on $\mathbf{Q}_L$:
\begin{equation}\label{PiOnQL}
    \hat{\mathcal{P}} =\left(
\begin{array}{ccccccc}
24-720 z_1 & 18 & -9-720 z_4 & -57-720 z_2\\
  8  &  2    &  -16 & -46+720 z_4 \\
  1  &  -2   &  -11  & -26\\
  -3   &  0    &  0   & -3+720 z_3 \\
\end{array}
 \right)
 \end{equation}\\
\\
\verb"Sixth step:"\\
Compute $(z_1,z_2,z_3,z_4)$ by requiring $\hat{\mathcal{P}}\equiv
\mathcal{H}$:
\begin{align*}
\begin{cases}
24-720 z_1 &=27\\
-57-720 z_2 &=-54\\
 -3+720 z_3 &=0\\
-9-720 z_4 &= -9\\
\end{cases}
\end{align*}
which gives:
\begin{align}\label{coefficienti}
\begin{cases}
z_1 &=-1/240\\
z_2 &=-1/240\\
z_3 &=1/240\\
z_4 &= 0\\
\end{cases}
\end{align}
Substituting \eqref{coefficienti} back into
\eqref{WantedInterpolating}, the final result is obtained:
\begin{align}\label{FinalResult}
\nonumber P(x,y) &=-3+\frac{69}{20}x+\frac{59}{60}x^2-
\frac{31}{16}x^3+\frac{25}{48}x^4-\frac{1}{80}x^5
-\frac{1}{240}x^6+\frac{103}{20}y+\\
\nonumber &-\frac{533}{60}xy-\frac{7}{16}x^2 y+\frac{13}{12}x^3
y-\frac{7}{16} x^4 y +\frac{1}{40}x^5 y
-\frac{119}{60}y^2+\\
\nonumber &+\frac{95}{16}x y^2-3 x^2 y^2+\frac{1}{8}x^3 y^2+
\frac{1}{16}x^4 y^2+\frac{7}{16}y^3-\frac{7}{6}x
y^3+\frac{7}{8}x^2 y^3+\\
&+\frac{1}{12}x^3 y^3+\frac{23}{48}y^4-\frac{1}{16}x
y^4-\frac{1}{16}x^2 y^4-\frac{7}{80}y^5+\frac{1}{40}x
y^5+\frac{1}{240}y^6
\end{align}
Note the difference between \eqref{FinalResult} and
\eqref{phi4interpol} although they interpolate the very same
matrix \eqref{Interpol}. The degree of $P$ is $6\equiv 2(L-1)$ as
stated in Theorem \ref{main}.

This procedure can be iterated without difficulties up to
interpolating inner-harmonic matrices of any size through a
repeated application of Lemma \ref{interpolation}. \vspace{14pt}

Hereafter we shall provide several preliminary lemmas which are
essential for the proof of the main result and have been hinted
previously.
\begin{Lemma}\label{basis}
Let $k>0$. Then $\mathbb{D}^\star_k$ is a vector space of dimension
$2$.
\end{Lemma}
First, we easily prove the following statement: let
$\mathbb{P}_{N}$ be the set of two variables polynomials up to
degree $N$ and let $P_1(x,y)\in\mathbb{P}_{N}$. Then
$P_2(x,y)=\dlap P_1(x,y)\in\mathbb{P}_{N-2}$.

In fact, we notice that the following properties hold:
\begin{align}
 \label{prop1}\dlap (ax^n+by^m) &= a\dlap
(x^n)+b\dlap (y^m) && \text{Linearity}\\
\label{prop2}\dlap (x^n y^m) &= x^n\dlap (y^m)+y^m\dlap (x^n) &&
\text{Leibniz rule}
 \end{align}

Furthermore, for every one-variable monomial in $x$ (or $y$), it is
straightforward to prove the following:
\begin{equation}
\dlap x^n=
\begin{cases}
-2\sum_{k=0}^{(n-2)/2}{n\choose 2k}x^{2k} &\text{if n
is even}\\
-2\sum_{k=0}^{(n-3)/2}{n\choose 2k+1}x^{2k+1} &\text{if n is odd}
\end{cases}
\end{equation}

Therefore, applying the Laplace operator to a one-variable
monomial of degree $n$, we obtain a linear combination of
one-variables monomials up to degree $n-2$. Thanks to
\eqref{prop1} and \eqref{prop2}, we can conclude that the same
holds also for two-variables polynomials.[QED]

It is well-known that $\mathbb{P}_{N}$ is a linear vector space,
with $\dim(\mathbb{P}_N)=\sum_{n=0}^N (n+1)=\frac{(N+1)(N+2)}{2}$ .
According to the previous results, we call
$\Pi_N:\mathbb{P}_{N}\rightarrow \mathbb{P}_{N-2}$ the following
linear map:
\begin{equation}
\Pi_N (P(x,y))=(\dlap P)(x,y)
\end{equation}
Then, we call $\mathbb{D}_N=\ker(\Pi_N)$, i.e. the following vector
subspace of $\mathbb{P}_N$:
\begin{equation}
\mathbb{D}_N=\{P(x,y)\in \mathbb{P}_N:\quad (\dlap
P)(x,y)=0\quad\forall (x,y)\in\mathbb{R}^2\}
\end{equation}
The elements of $\mathbb{D}_N$ are discrete harmonic polynomials.
The dimension of $\mathbb{D}_N$ can be found simply applying the
Rank-nullity theorem to the map $\Pi_N$:
\begin{equation}
\dim(\mathbb{D}_N)=\dim(\mathbb{P}_N)-\dim(\mathbb{P}_{N-2})=\frac{(N+1)(N+2)}{2}-\frac{N(N-1)}{2}=
2N+1
\end{equation}
Let $\mathbb{D}_k^\star$ be the following vector subspace of
$\mathbb{D}_N$:
\begin{equation}
\mathbb{D}_k^\star=\{P(x,y)\in \mathbb{D}_N|\quad \text{P's degree
is exactly }k\leq N\}
\end{equation}
Obviously,
$\dim(\mathbb{D}_k^\star)=\dim(\mathbb{D}_k)-\dim(\mathbb{D}_{k-1})=2$.

Therefore, for $k>0$ we can always find two (and not more) linearly
independent discrete harmonic polynomials, i.e. elements of
$\mathbb{D}_N$, with the same degree $k$.

Following the standard algebraic procedure, it is quite easy to
build up a complete basis
$\mathcal{B}^{\star}=\{e^{\star}_{1},\ldots,e^{\star}_{2N+1}\}$ for
$\mathbb{D}_N$, starting from the canonical basis in $\mathbb{P}_N$:
\begin{equation}\mathcal{B}_N=\{1,x,y,x^2,xy,y^2,\ldots,y^N\} \nonumber\end{equation}

Throughout this paper, we will refer to the basis $\{U_k(x,y)\}$
listed in the Appendix.

\begin{Lemma}\label{unicitycompletion}
For every square matrix with an arbitrary fixed rational contour,
there exists one and only one inner-harmonic completion.
\end{Lemma}

Let $F_{\Omega}(i,j):\mathbf{Q}_L\rightarrow \mathbb{Q}$ and let its
$(4L-4)$ border sites be forced to assume rational values $z_k$
belonging to the set $\Omega$.

In matrix form, we have:
\begin{equation}
F_{\Omega}=\left(
\begin{array}{cccccc}
  z_1  &  z_2 &  z_3 & \cdots & \cdots &  z_L\\
  z_{4L-4}  &  x_1 & x_2  &  \cdots  &  x_{L-2} & z_{L+1}\\
  z_{4L-5}  & x_{L-1}  &  x_{L} & \cdots  &  x_{2(L-2)}  &  z_{L+2}\\
  \vdots  &  \vdots  &  \vdots  &  \vdots & \ddots & \vdots\\
  z_{3L-2} & z_{3L-3} & \cdots & \cdots & z_{2L} &
  z_{2L-1}\\
\end{array}
 \right)
\end{equation}

The nested $(L-2)\times (L-2)$ submatrix $F^\dag_\Omega$ has unknown
entries $x_j\in {\mathbb{Q}}$.

We prove that, for each set $\Omega$, there exists one and only one
submatrix $F^\dag_\Omega$ with rational entries such that
$F_{\Omega}(i,j)$ is inner-harmonic.

If we impose the inner-harmonicity condition on $F_{\Omega}$, we get
the linear system $\mathbf{\hat{A}}\vec{x}=\vec{\eta}(\{z_k\})$,
where $\mathbf{\hat{A}}$ is the following $(L-2)^2\times (L-2)^2$
matrix:
\begin{equation}\label{forb2}
\mathbf{\hat{A}}=4\mathbf{\hat{I}
}-{\mathbf{\hat{H}}}=4\mathbf{\hat{I}}-\left(
\begin{array}{cccccc}
  {\mathbf{H}}  &  \mathbf{I} &  \mathbf{0} &  \cdots & \mathbf{0} & \mathbf{0}\\
  \mathbf{I}  &  {\mathbf{H}}  &  \mathbf{I}  &  \cdots & \cdots & \mathbf{0}\\
  \mathbf{0}  &  \mathbf{I}  &  {\mathbf{H}}  &  \mathbf{I} & \cdots & \mathbf{0} \\
  \vdots  &  \vdots  &  \vdots  &  \vdots & \ddots & \vdots\\
 \vdots  &  \vdots  &  \vdots  &  \vdots & \ddots &  \mathbf{I} \\
  \mathbf{0} & \cdots & \cdots & \cdots & \mathbf{I} &
  {\mathbf{H}}\\
\end{array}
 \right)
\end{equation}
$\mathbf{\hat{I}}$ is the identity matrix $(L-2)^2\times (L-2)^2$,
$\mathbf{I}$ is the identity matrix $(L-2)\times (L-2)$,
$\mathbf{0}$ is the null matrix $(L-2)\times (L-2)$ and
${\mathbf{H}}$ is a well-known matrix describing the Hamiltonian of
nearest-neighbor hopping on a one-dimensional lattice (see \cite{MM}
and references therein):
\begin{equation}\label{forb3}
{\mathbf{H}}=\left(
\begin{array}{cccccc}
  0  &  1 &  0 &  0 & \cdots & 0\\
  1 &  0  &  1  &  0 & \cdots & 0\\
  0 &  1  &  0  &  1 & \cdots & 0 \\
  \vdots  &  \vdots  &  \vdots  &  \vdots & \ddots & \vdots\\
 \vdots  &  \vdots  &  \vdots  &  \vdots & \ddots & 1\\
  0 & 0 & 0 & \cdots & 1& 0\\
\end{array}
 \right)
\end{equation}

The vector $\vec{\eta}$ depends on the fixed contour values. Its
entries are of the following forms:
\[\eta_j=
\begin{cases}
z_{\alpha}+z_{\beta} & \text{if $x_j$ is a corner site of
$F^\dag_\Omega$}\\z_{\gamma} & \text{if $x_j$ is a border site of
$F^\dag_\Omega$, but not a corner site}\\0 & \text{otherwise}
\end{cases}\]


Since the matrix \eqref{forb2} is diagonal predominant
\cite{Golub}, the system admits one and only one solution in
$\mathbb{Q}$. [QED]

\begin{Corollary}\label{corollary}
If $\Omega=\{0,\ldots,0\}$, then $F_{\Omega}$ is the null matrix
$L\times L$.
\end{Corollary}

\begin{Lemma}\label{3x3}
Given a $3\times 3$ inner-harmonic matrix $\mathcal{A}$, it is
always possible to find a discrete harmonic polynomial $P(x,y)$
with rational coefficients and degree $4$ such that
$P\doteq\mathcal{A}$ on $\mathbf{Q}_3$.
\end{Lemma}

For $L=3$, there are $8$ sites along the contour. Choose the
following set of discrete harmonic polynomials\footnote{Obviously,
infinitely many other choices are equally possible.} (see
Appendix):
\begin{equation}\label{ChooseSet}
  \{U_0(x,y),\ldots,U_6(x,y)\}\cup \{U_8(x,y)\}
\end{equation}
The sought polynomial $P(x,y)$ satisfying the Lemma may be written
as a linear combination of the polynomials in \eqref{ChooseSet},
with unknown coefficients $\alpha_j$ ($j=1,\ldots,8$).

The condition that $P\doteq\mathcal{A}$ translates into a linear
system with $8$ equations in the unknowns $\alpha_j$, whose matrix
of coefficient for the choice \eqref{ChooseSet} is:
\begin{equation}\label{MatrixOfCoefficients}
{\mathbf{M}}=\left(
\begin{array}{cccccccc}
  1  &  0 &  0 &  0 & 0 & 0 & 0 & 0\\
  1  &  0 &  1 &  0 & 1 & 0 & 1 & -1\\
  1  &  0 &  2 &  0 & 4 & 0 & 8 & 8\\
  1  &  1 &  0 &  0 & -1 & 1 & 0 & 1\\
  1  &  1 &  2 &  2 & 3 & -11 & 2 & -15\\
  1  &  2 &  0 &  0 & -4 & 8 & 0 & 16\\
  1  &  2 &  1 &  2 & -3 & 2 & -11 & -9\\
  1  &  2 &  2 &  4 & 0 & -16 & -16 & -72\\
\end{array}
 \right)
\end{equation}
The determinant of $\mathbf{M}$ is nonzero. Thus the polynomial
interpolating the contour (and for Lemma \ref{unicitycompletion}
also the central site) always exists and has degree $4$. [QED]

\begin{Lemma}\label{ZLpolynomials}
For every $L\geq 3$, there exist four discrete harmonic
\textsf{(L)-polynomials} $\xi_1$, $\xi_2$, $ \xi_3$ and $\xi_4$,
whose degree is less or equal to $2L$, such that $\xi_k\doteq
Z^{(k)},~k=1,2,3,4$. The entries of the matrices
$Z^{(k)},~k=1,2,3,4$ are all $0$ except:
\begin{enumerate}
\item the entry $(0,L)$ for $Z^{(1)}$;
\item the entry $(L,L)$ for $Z^{(2)}$;
\item the entry $(L,0)$ for $Z^{(3)}$;
\item the entries $(L-1,L)$ and $(L,L-1)$ for $Z^{(4)}$;
\end{enumerate}
\end{Lemma}

As it was evident from the example of application, the
\textsf{(L)-polynomials} have the following effect. Given a
$(L+1)\times (L+1)$ inner-harmonic matrix $\mathcal{G}$ and a
discrete harmonic polynomial $P(x,y)$ interpolating the lower-left
\emph{minor} ($L \times L$) of $\mathcal{G}$, those polynomials
neutralize the mismatch between $4$ sites along the border of
$\mathcal{G}$ and the values assumed by $P(x,y)$ on
$\mathbf{Q}_{L+1}$.

We prove now the existence of $\xi_1(x,y)$. For the others, the
procedure is completely analogue.

Consider a set of $4L$ linearly independent discrete harmonic
polynomials $P_{k,s}(x,y)$, where $k=1,...,2L$ is the degree, and
$s=1,2$ which do not contain the constant term.

We write the sought $\xi_1(x,y)$ in the form
\begin{equation}
\label{zerop} \xi_1(x,y) =
\alpha_1P_{1,1}(x,y)+\alpha_2 P_{1,2}(x,y)+...+\alpha_{4L}
P_{2L,2}(x,y)
\end{equation}

To determine the $4L$ unknowns $\alpha_j$, we require that
$\xi_1(x,y)$ should be zero on i) the border sites of the
lower-left minor $\mathcal{M}$ of $Z^{(1)}$ ii) four other points
in $\mathbb{Z}_2$, precisely: $(L-1,L),(L,L),(L,0),(L+1,L)$.

This translates into a linear homogeneous system $\mathcal{S}$ in
$4L$ equations for the $4L$ unknowns $\alpha_1,...,\alpha_{4L}$.
Note that the site $(L,L-1)$ is automatically zero due to the
harmonicity condition.

The first row of the matrix of coefficients for $\mathcal{S}$ is
given by:
\begin{equation} P_{1,1}(0,0), P_{1,2}(0,0),...,P_{2L,2}(0,0)
\end{equation}
and these values are all zero because the polynomials do not
contain the constant term.

Thus, the determinant is zero and the homogeneous system has an
infinite non-zero solutions set $\{\alpha_1,...,\alpha_{4L}\}$.
Since the polynomials $P_{k,s}(x,y)$ are linearly independent, the
obtained polynomial cannot be identically zero by definition.

Now, let $\xi_1$ be defined by a non-zero solution
$\{\alpha_1,...,\alpha_{4L}\}$. Being zero along the contour of
$\mathcal{M}$, it is zero inside $M$ because of the Corollary
\ref{corollary}.

It is also zero by the discrete harmonicity relation on sites
$(k,L),~ k=1,..,L+1$, and sites $(L,k),~k=0,...,L+1$. Instead, it
is required to be nonzero on the site $(0,L)$. Indeed, we can
prove that this is the case by contradiction. Assume that
$\xi_1(0,L)=0$. We have:
\begin{equation}\xi_1(j,k)=0,\quad k=L,\quad j=0,1,...,L+1~.\end{equation}
Let:
\begin{align}
n &=L/2+1 & m &=L/2 && \text{for even L}\\
n &=(L+1)/2+1 & m &=(L+1)/2-1 && \text{for odd L}
\end{align}

Due to the harmonicity relation, there is an integer $J$, $0<J<L+1$,
such that:\\
\begin{align*}
\xi_1(J,L-1+i) &=0 &&\text{for every } i=1,...,n \\
\xi_1(J,-k) &=0 &&\text{for every } k=1,...,m
\end{align*}

This means that the one variable polynomial $\eta(y)=\xi_1(J,y)$
has $2L+1$ zeros: but this is absurd, since its degree in $y$ is
at most $2L$. Therefore $\xi_1(0,L)\neq 0$.[QED]

\begin{Lemma}\label{interpolation}
Let $A$ be a inner-harmonic matrix of order $L$, and $A^\prime$
the $(L-1)\times(L-1)$ lower-left inner-harmonic minor of $A$. Let
$\chi(x,y)$ be a discrete harmonic polynomial of degree $h$
interpolating $A^\prime$. Then, it is possible to define a
discrete harmonic polynomial $\sigma (x,y)$, of degree $k= \max
[2(L-1),~h]$, interpolating $A$.
\end{Lemma}
Define:
\begin{align*}
\begin{cases}
s_1 &:= \text{Site $(0,L-1)$}\\
s_2 &:= \text{Site $(L-2,L-1)$}\\
s_3 &:= \text{Site $(L-1,L-1)$}\\
s_4 &:= \text{Site $(L-1,L-2)$}\\
s_5 &:= \text{Site $(L-1,0)$}\\
\end{cases}
\end{align*}
and denote $\chi(s_k):=\chi_k$ and $A(s_k):=a_k$ for simplicity.

We write the sought $\sigma (x,y)$ in the form:
\begin{equation} \label{sigma} \sigma(x,y)=
\chi(x,y)+\sum_{k=1}^4z_k \xi_k(x,y)~,
\end{equation}
where the $\xi_k(x,y)$ are \textsf{(L-1)-polynomials} as defined
in Lemma \ref{ZLpolynomials}, and $z_k$ are coefficients to be
determined. The degree of each of the $\xi_k$ is at most $2(L-1)$,
confirming the statement of the Lemma about the degree of
$\sigma$.

We note that $\sigma(x,y)\equiv\chi(x,y)$ on the sites of
$A^\prime$, since all the \textsf{(L-1)-polynomials} assume value
0 there.

The values assumed by the polynomial $\chi$ on the North and East
borders of $\mathbf{Q}_L$ are uniquely constrained by the
harmonicity condition, except the five sites $s_k$. In general,
$a_k\neq \chi_k$.

For example, for $L=5$ we have the following schematic situation
(compare with \eqref{Hhat4}):
\begin{equation}\label{Situation}
A=\left(
\begin{array}{ccccc}
  \square  &  \blacksquare &  \blacksquare &  \lozenge & \square \\
  \cdot  &  \cdot &  \cdot &  \cdot & \lozenge\\
  \cdot  &  \cdot &  \cdot &  \cdot & \blacksquare\\
 \cdot  &  \cdot &  \cdot &  \cdot & \blacksquare\\
 \cdot  &  \cdot &  \cdot &  \cdot & \square\\
\end{array}
 \right)
\end{equation}
where:
\begin{align*}
\begin{cases}
\cdot &\Rightarrow \text{Sites in $A^\prime$, where $\chi\doteq A$}\\
\blacksquare & \Rightarrow\text{Sites where $\chi\doteq A$ by harmonicity}\\
\square & \Rightarrow\text{Sites ($s_1,s_3,s_5$) where $\chi\cancel{\doteq} A$ }\\
\lozenge &\Rightarrow \text{Sites ($s_2,s_4$) where $\chi\cancel{\doteq} A$, but mutually constrained by harmonicity}\\
\end{cases}
\end{align*}
Indeed, the discrete harmonicity condition, applied to $A$ and
$\chi$, requires that:
\begin{equation}\label{harmcond}
\chi_2+\chi_4=a_2+a_4
\end{equation}

Given that $\xi_k(s_1)=0$ for $k\neq 1$ and $\xi_1(s_1)=\gamma\neq
0$ (Lemma \ref{ZLpolynomials}), we get from equation
(\ref{sigma}):
\begin{equation}
\sigma(s_1)= \chi_1 +z_1\gamma=a_1
\end{equation}
This determines $z_1$ as $z_1= (a_1-\chi_1)/\gamma$.

The same procedure applies to the sites $s_3$ and $s_5$,
determining $z_2$ and $z_3$: note the shift of indices, reflecting
the fact that we have five sites and only four
\textsf{(L-1)-polynomials}.

In fact, the polynomial $\xi_4$ has to be nonzero simultaneously
on both sites $s_2$ and $s_4$, and by harmonicity $\xi_4(s_2) =
-\xi_4(s_4)$. This constraint, however, is compatible with the
correct definition of $z_4$ and therefore of $\sigma(x,y)$.

Indeed, define $\omega =\xi_4(s_2)=-\xi_4(s_4)$. Equation
(\ref{sigma}) requires evidently that
$\sigma(s_2)=\chi_2+z_4\omega=a_2$ and
$\sigma(s_4)=\chi_4-z_4\omega=a_4$. Both equations are obviously
satisfied by $z_4=(a_2-\chi_2)/\omega$ thanks to \eqref{harmcond}.

Thus, the coefficients $z_1,..,z_4$ in (\ref{sigma}) are uniquely
determined and the polynomial $\sigma(x,y)$ interpolating
$A$ exists. [QED]\\
\newline
We are now able to provide a proof of Theorem \ref{main}.

We only need an iterative (or ``telescopic'') application of
previous results: starting from $\mathcal{H}$, we drop the upper
row and last column on the right, defining the minor
$\mathcal{H}^{(1)}$.

If we can find a discrete harmonic polynomial
$\chi(x,y)\doteq\mathcal{H}^{(1)}$, such that
$\mathrm{deg}(\chi)\leq 2(L-1)$, the Theorem follows via Lemma
\ref{interpolation}; otherwise, we drop the upper row and last
column on the right of $\mathcal{H}^{(1)}$ again, and restart the
procedure.

This process is consistent, because the minors iteratively defined
continue to be inner-harmonic.

Suppose that we have finally found the minor $\mathcal{H}^{(n)}$
(whose size is $L-n$) of $\mathcal{H}^{(n-1)}$, admitting an
interpolating polynomial $\chi_{(n)}(x,y)$ such that
$\mathrm{deg}(\chi_{(n)})\leq 2(L-1)$. By Lemma
\ref{interpolation}, the minor $\mathcal{H}^{(n-1)}$ can be
interpolated, and so on, up to interpolating $\mathcal{H}$.

Since at least for $L=3$ the interpolating polynomial always
exists (Lemma \ref{3x3}), in the worst possible case the
telescopic algorithm will start from the $3\times 3$ lower left
minor of $\mathcal{H}$ , and will eventually produce the desired
result by repeated applications of Lemma \ref{interpolation}.
[QED]
\section{Final remarks}\label{FinalRemarks}
In this note, we have developed a ``telescopic'' technique to
interpolate an inner-harmonic matrix of size $L$ by a discrete
harmonic polynomial of degree less or equal to $2(L-1)$.

The solution we have presented proves a conjecture about hidden
conservation laws in the context of some statistical mechanics
models, namely the so called fixed-energy sandpiles with
deterministic \textsf{BTW} toppling rule.

We remark that the algorithmic procedure we devised should be
regarded as a mere tool for the proof, and by no means is meant to
provide a computationally efficient and robust interpolator for
inner-harmonic matrices.

As a final point, we wish to give here a short survey on other
related questions and problems which have not been addressed in
this paper and could be worthy of further investigations.
\begin{enumerate}
  \item \emph{Discrete harmonic polynomials of minimal degree:} the
  constructive procedure outlined in section \ref{Main Results} does
  not lead to an uniquely defined interpolating polynomial. A
  natural question to ask is what the minimal attainable degree of
  such a polynomial is, and how to build it up.
  \item \emph{A related combinatorial problem:}
  Another class of matrices ($\mathcal{M}_L^\star$) with \emph{integer} entries and
  closely related to $\mathcal{M}_L$ emerges in \cite{Sandpile}
  and proves to be connected to deep symmetries of the evolving
  rule of that model.\\The main features of $\mathcal{M}_L^\star$ are:
\begin{itemize}
  \item Condition \eqref{InnHarmMatCondition} holds \emph{modulus} the size
  $L$ of the matrix.
  \item Cyclical border conditions are imposed and condition
  \eqref{InnHarmMatCondition}
  holds for border sites as well.
  \item Entries are bounded by an integer $M$.
\end{itemize}
An interesting problem in analytical combinatorics, with many
possible consequences on the underlying physical issue, is to
count the number of those matrices for fixed $L$ and $M$.
\end{enumerate}

\begin{acknowledgements}
We thank Dr. Igor Krasovsky, Dr. Ilia Krasikov and Dr. Steven
Noble (Brunel University) for helpful comments. We also thank
Elisa Garimberti (Brunel University) for a careful revision of the
manuscript.
\end{acknowledgements}

\section*{Appendix: list of Discrete Harmonic Polynomials}
We report here a basis of discrete harmonic polynomials up to
degree $9$ that we used repeatedly throughout the paper:
\begin{align}
U_0(x,y) &= 1\\
U_1(x,y) &= y\\
U_2(x,y) &= x\\
U_3(x,y) &=xy\\
U_4(x,y) &=x^2-y^2\\
U_5(x,y) &=-3 x^2 y+y^3\\
U_6(x,y) &=x^3-3 x y^2\\
U_7(x,y) &=x^3 y-x y^3\\
U_8(x,y) &=x^4-2 x^2-6 x^2 y^2+y^4\\
U_9(x,y) &= 5 x^4 y-10 x^2 y^3-10 x^2 y+y^5\\
U_{10}(x,y) &= x^5-10 x^3 y^2+5 x y^4-10 x y^2\\
U_{11}(x,y) &= x^5 y-\frac{10}{3}x^3 y^3-\frac{10}{3} x y^3 +x y^5\\
U_{12}(x,y) &=-15 x^4 y^2 - 10 x^4 + 10 x^2 + 15 x^2 y^4 + 30 x^2 y^2 - y^6 + x^6\\
U_{13}(x,y) &=35 x^4 y^3+70 x^4 y-21 x^2 y^5-70 x^2 y^3-70 x^2 y+y^7-7 x^6 y\\
U_{14}(x,y) &=-21 x^5 y^2 - 70 x^3 y^2 + 35 x^3 y^4 - 7 x y^6 + 70 x y^4 - 70 x y^2 + x^7\\
U_{15}(x,y) &= -7 x^5 y^3 + 7 x^3 y^5-\frac{70}{3}x^3
y^3-\frac{70}{3}x y^3
+x y^7+14 x  y^5 + x^7 y\\
\nonumber U_{16}(x,y) &= -140 x^4 y^2 + 70 x^4 y^4 - 140 x^4 + 166
x^2 - 28
x^2 y^6 + 280 x^2 y^4 +\\
  &+560 x^2 y^2 + y^8 - 28y^6 + x^8 - 28 x^6 y^2\\
\nonumber U_{17}(x,y) &= 126 x^5 y^4 - 252 x^5 y^2 - 84 x^3 y^6 -
840 x^3 y^2 + 840 x^3 y^4 +
  9 x y^8 +\\
  &- 252x y^6 + 1260 x y^4 - 1026 x y^2 + x^9 - 36 x^7 y^2\\
\nonumber U_{18}(x,y) &= 840 x^4 y^3 + 126 x^4 y^5 + 1260 x^4 y -
252 x^2 y^5 - 36 x^2 y^7 -
  840 x^2 y^3 +\\
  &- 1026 x^2 y + y^9 + 9 x^8 y - 84 x^6 y^3 - 252 x^6  y
\end{align}

\end{article}
\end{document}